\documentclass[prd,twocolumn,amsmath,amssymb,floatfix,superscriptaddress,notitlepage,nofootinbib]{revtex4-1}

\usepackage{amsfonts,amssymb,amsmath,graphicx,color,bm}
\usepackage{enumitem}
\definecolor{ultramarine}{rgb}{0.07, 0.04, 0.56}
\definecolor{cadmiumgreen}{rgb}{0.0, 0.42, 0.24}
\definecolor{indigo(dye)}{rgb}{0.0, 0.25, 0.42}
\usepackage[linktocpage=true]{hyperref}
\hypersetup{
colorlinks=true,
citecolor=ultramarine,
linkcolor=cadmiumgreen,
urlcolor=indigo(dye),
}

\newcommand{\f}[2]{\frac{#1}{#2}}  
\newcommand{\mk}[1]{\left( #1 \right)}  
\newcommand{\kk}[1]{\left[ #1 \right]}  
  
\newcommand{\be}{\begin{equation}}  
\newcommand{\ee}{\end{equation}}
\newcommand{\bem}{\begin{pmatrix}}
\newcommand{\eem}{\end{pmatrix}}

\newcommand{\Ca}{C_1}
\newcommand{\Cb}{C_2}
\newcommand{\Cc}{C_3}
\newcommand{\Ns}{N_s}

\begin{document}

\title{
Primordial Black Holes and Slow-Roll Violation
}

\author{Hayato Motohashi}
\affiliation{Instituto de F\'{i}sica Corpuscular (IFIC), Universidad de Valencia-CSIC,
E-46980, Valencia, Spain}

\author{Wayne Hu}
\affiliation{Kavli Institute for Cosmological Physics, Department of Astronomy and Astrophysics, The University of Chicago, Chicago, Illinois 60637, USA}

\begin{abstract}
For primordial black holes (PBH) to be the dark matter in single-field inflation, the slow-roll approximation must be violated by at least ${\cal O}(1)$
in order to enhance the curvature power spectrum within the required number of efolds 
between CMB scales and PBH mass scales.
Power spectrum predictions which rely on the inflaton remaining
on the slow-roll attractor can fail dramatically leading to qualitatively incorrect conclusions in models like an inflection potential and misestimate the mass scale in a running mass model.
We show that an optimized temporal evaluation of the Hubble slow-roll parameters to 
second order remains a good description  for a wide range of PBH
formation models where up to a $10^7$ amplification of power occurs in $10$ efolds or more.
\end{abstract}


\maketitle  


\section{Introduction}

With the detection of black hole mergers by LIGO~\cite{Abbott:2016blz,Abbott:2016nmj,Abbott:2017vtc} and the continuing lack of direct detection of WIMPs, interest in primordial black holes (PBHs)~\cite{1966AZh....43..758Z,Hawking:1971ei,Carr:1975qj} as 
candidates for dark matter has recently received renewed attention \cite{Bird:2016dcv,Clesse:2016vqa,Sasaki:2016jop,Kawasaki:2016pql,Carr:2016drx,Inomata:2016rbd,Nakama:2016gzw,Kuhnel:2017pwq,Chiba:2017rvs,Carr:2017jsz}.
PBHs form from density fluctuations that are comparable to or exceed order unity at horizon 
crossing.   For PBHs to be the dark matter therefore requires a large amplification of the
inflationary  power spectrum between cosmic microwave background (CMB) scales and PBH mass scales.

In \S\ref{sec:nogo}, we show that independently of the detailed model of inflation, this amplification
requires at least an ${\cal O}(1)$ violation of slow roll in canonical single-field inflation.
In the recent literature, this has lead to confusion as to which inflaton potentials support PBH formation~\cite{Garcia-Bellido:2017mdw,Ezquiaga:2017fvi} and how best to
calculate the power spectrum for cases that do \cite{Drees:2011hb}.

Conversely a much larger than ${\cal O}(1)$ violation is not required and so the various 
slow-roll (SR) and optimized slow-roll (OSR) approximations for the power spectrum  reviewed in Appendix~\ref{sec:D2fs} 
perform differently.   Using the inflection (\S\ref{sec:im}), running mass (\S\ref{sec:rm}) and 
slow-roll step (\S\ref{sec:eps}) models as examples, we show that the potential slow-roll formalism
can yield qualitatively incorrect results whereas the optimized Hubble slow-roll hierarchy~\cite{Motohashi:2015hpa,Motohashi:2017gqb}
can provide accurate results in most of the cases relevant to PBH formation where
slow roll is not grossly violated.

\section{No go for slow roll}
\label{sec:nogo}

Following the pioneering work of \cite{Carr:1975qj}, we model PBH formation using the  Press-Schechter (PS) approach
where the PBH  fraction is determined by a
collapse threshold $\delta_c$ in the density field smoothed with a Gaussian at the horizon reentry scale.
In order to relate the collapse fraction directly to the curvature power spectrum, we approximate the
result with the analogous probability of Gaussian curvature fluctuations lying above a threshold $\zeta_c$~\cite{Drees:2011hb,Harada:2013epa,Young:2014ana} 
\begin{align} \label{betaz}
\beta &\equiv \f{\rho_{\rm PBH}}{\rho_{\rm tot}} 
= 2 \int^\infty_{\zeta_c} d\zeta\, \f{1}{\sqrt{2\pi }\Delta_\zeta} e^{-\zeta^2/(2\Delta_\zeta^2)} \notag\\
&= {\rm erfc} \mk{\f{\zeta_c}{\sqrt{2} \Delta_\zeta}} 
\approx \sqrt{\f{2}{\pi }}\, \f{\Delta_\zeta}{\zeta_c} e^{-\zeta_c^2/(2\Delta_\zeta^2)} ,
\end{align}
where the approximation assumes $\zeta_c \gg \Delta_\zeta$ so that PBHs form from
rare peaks.   Here the density and curvature thresholds are related by assuming a nearly scale-invariant
curvature power spectrum for a few efolds around horizon crossing 
\cite{Drees:2011hb,Young:2014ana}
\be \zeta_c = \f{9}{2\sqrt{2}} \delta_c . \ee
In the following estimates, we take $\zeta_c=1.3$ 
based on studies that show a range
of $\delta_c = 0.4-0.5$ \cite{Musco:2012au,Harada:2013epa}.
The additional factor 2 in (\ref{betaz}) is the usual PS bookkeeping that accounts for locally under threshold regions collapsing in globally over threshold regions.
Note that $\Delta_\zeta^2$ is the variance of the curvature field per logarithmic interval
in $k$ and hence $\beta$ represents the collapse fraction per $d\ln k$ in the spectrum.

A large collapse fraction requires a large amplitude of $\Delta_\zeta^2$, much larger than 
\begin{equation}
\Delta_\zeta^2(k_0) \approx 2.2\times 10^{-9}
\label{CMBnorm}
\end{equation}  
measured by the CMB at $k_0 = 0.05$ Mpc$^{-1}$~\cite{Ade:2015xua}.
Since the leading-order SR approximation gives the curvature power spectrum as
\begin{equation}
\Delta_\zeta^2 = \frac{ H_I^2 }{8\pi^2 \epsilon_H},
\label{SRpower}
\end{equation} 
where $\epsilon_H = -d\ln H_I/dN$ with $H_I$ as the Hubble parameter during inflation and
$N$ as the efolds of inflation, the amplification of the power spectrum required for PBH formation can in principle be achieved by making $\epsilon_H$ small.

One might think  that the slow-roll approximation still holds so long as $\epsilon_H$ is small and 
the expansion is nearly de Sitter $H_I \approx$\,const. 
However, even if $\epsilon_H$ itself is small, if its fractional variation per efold violates 
\begin{equation}
\left| \frac{\Delta \ln \epsilon_H}{\Delta N}\right|  \ll 1
\label{eqn:slowrollviolation}
\end{equation}
then the slow-roll condition is broken.  
The various slow-roll approximations for the power spectrum reviewed in Appendix~\ref{sec:D2fs}
all fail if (\ref{eqn:slowrollviolation}) is strongly violated for a sustained period and perform differently for ${\cal O}(1)$ or transient violation 
~\cite{Kinney:2005vj,Dvorkin:2009ne,Namjoo:2012aa,Martin:2012pe,Motohashi:2014ppa,Motohashi:2017aob,Motohashi:2017vdc}.
Indeed without specifying a specific inflationary model, we shall now show that for PBHs to be all of the dark matter 
there must be at least an ${\cal O}(1)$ violation of the slow-roll condition (\ref{eqn:slowrollviolation})
in single-field inflation.

The collapse fraction required for all of the dark matter to be PBHs depends on their mass.   
Smaller masses enter the horizon earlier during radiation domination and redshift more slowly
than radiation once they collapse.   
We define the PBH mass  as $M =\gamma M_H$, where 
\begin{equation}
M_H\equiv \f{4\pi \rho}{3H^3}=\f{1}{2GH}
\end{equation}
is the horizon mass and $\gamma$ accounts for the efficiency of collapse.  
During radiation domination, the Hubble parameter $H$ is given by
\be \label{FeqRD} H^2 = \left( \frac{g_{*}}{g_{*0}}\right)^{-1/3} \Omega_r H_0^2 a^{-4}, \ee
where we have assumed that the effective degrees 
of freedom in the entropy and energy densities approximately coincide
$g_{*S}\approx g_*$ as they do before electron-positron annihilation in the standard model.  Here
$g_{*0}=3.36$ is the value of the latter today and the radiation energy density today is given by
$\Omega_r h^2 = 4.18\times 10^{-5}$ for $T_{\rm CMB}=2.725\,$K and $N_{\rm eff} = 3.046$.
We then obtain $ M$ as a function of the horizon crossing epoch $a_H$
\be  M = \frac{\gamma}{2 GH} = 4.84 \times 10^{24} \gamma \left( \frac{g_*}{g_{*0}}\right)^{1/6} a_H^2 M_\odot .\ee

After formation, the PBH density dilutes as matter so the relic density today in
a mass range around $ M$ of $|d\ln M| = 2 d\ln k$ is given by
\be \frac{d\Omega_{\rm PBH} h^2}{d\ln M}  =\frac{1}{2} \beta(M)\left( \frac{g_*}{g_{*0}} \right)^{-1/3} \Omega_r h^2 a_H^{-1},\ee
where we ignore mass accretion and evaporation. 
If we define
\be \bar \beta= \frac{a_H}{2}\int_M^\infty  \frac{dM'}{M'}\frac{\beta}{a_H} ,\ee
we obtain the cumulative abundance $>M$
\be \label{PBHdensity}
\Omega_{\rm PBH} h^2  = \bar \beta \left( \frac{g_*}{g_{*0}} \right)^{-1/3} \Omega_r h^2 a_H^{-1}, \ee
Note that if $\beta(M)=$\,const.\  as it is for a scale-invariant power spectrum, then $\bar \beta=\beta$.

We can invert (\ref{PBHdensity}) to find the value of 
$\bar\beta$ required to produce a given relic density 
\be \label{betam} \bar \beta = 1.3 \times 10^{-9} \gamma^{-1/2} \left(  \frac{g_*}{g_{*0}} \right)^{1/4}
\left( \frac{\Omega_{\rm PBH} h^2}{0.12} \right) \left(\frac{ M}{M_\odot} \right)^{1/2} . \ee
Given that the cold dark matter density $\Omega_c h^2 = 0.1199\pm 0.0022$~\cite{Ade:2015xua}, this gives the $\bar\beta$ required for all of the dark matter to be in PBHs above $M$.
We can then set \eqref{betam} equal to \eqref{betaz} to obtain necessary local
value of $\Delta_\zeta^2$ corresponding to the chosen mass $M$.

In order to determine whether the slow-roll condition (\ref{eqn:slowrollviolation}) is violated, we next need to estimate the change in efolds $\Delta N$
during inflation over which this enhancement occurs.   
The  comoving scale corresponding to $M$ is $a_H H=a_{\rm exit} H_I$
and it exited the horizon during inflation $N$ efolds after the CMB
scale $k_0=0.05$\,Mpc$^{-1} = a_0 H_I$ exited. 
Assuming that $H_I\approx$ const., we can estimate the efolds by the ratio of comoving scales 
\begin{align} \label{Nm} 
N &=\ln \left( \frac{a_{\rm exit}}{a_0} \right)= \ln \left( \frac{a_H H}{0.05\, {\rm Mpc}^{-1} }\right) \notag\\
&= 18.4 - \frac{1}{12} \ln \frac{g_*}{g_{*0}} + \frac{1}{2}\ln \gamma -\frac{1}{2} \ln \frac{ M}{M_\odot}.
\end{align}
This  sets an upper bound on the duration of the change $\Delta N\le N$.

To establish a lower bound of the variation $|\Delta \ln \epsilon_H/\Delta N|$, let us consider the
largest $\Delta N$ or the smallest PBH mass that can compose the dark matter.
Low mass PBHs evaporate by Hawking radiation.   Even if the mass of PBHs grows after formation by merging and accretion, in order to be the dark matter they must at least
survive until  matter radiation equality.  We therefore equate
the time scale for evaporation by Hawking radiation 
\be t_{\rm ev} = \f{5120\pi G^2M^3}{\hbar c^4} = 6.6 \times 10^{74} \mk{\f{M}{M_\odot}}^3 ~{\rm s} \ee
to the time of equality using the radiation dominated estimate of \eqref{FeqRD} and
$t \approx 1/2H$ to obtain
\be \label{Mevap} M_{\rm min} 
= 1.5\times 10^{-21} \left(\frac{\Omega_m h^2}{0.14}\right)^{-2/3} M_\odot .\ee
With the Planck constraint $\Omega_m h^2 = 0.1426 \pm 0.0020$~\cite{Ade:2015xua},
this scale left the horizon at
\be N \approx 42 +\frac{1}{2} \ln \gamma \ee
relative to CMB scales where we have used $g_*=106.75$ which
is appropriate for  $M<1.2\times 10^{-6} \gamma M_\odot$ in the standard model.

We can now put this together to place a lower bound on the level of slow-roll violation. 
Since $\gamma<1$ requires  larger $\beta$ and larger $\Delta^2_\zeta$, we set 
$\gamma=1$ to be conservative and from  \eqref{betaz} and \eqref{betam} obtain 
\be \label{minenh} \Delta_\zeta^2 = 0.021, \ee
as the level of locally scale-invariant power ($\bar\beta=\beta$) required at this mass.
Given $\Delta_\zeta^2(k_0)$ on CMB scales (\ref{CMBnorm}),
\be\left|  \f{\Delta\ln\epsilon_H}{\Delta N} \right| > 0.38 . \ee
We conclude that for PBHs to be the dark matter in single-field inflation, the slow-roll condition must be
violated at least at ${\cal O}(1)$.

Although the various approximations employed in this bound carry large uncertainties, including the choice of $\gamma$, the Gaussian approximation for rare peaks, etc., they all enter into
$\Delta\ln \epsilon_H/\Delta N$ logarithmically.   Even orders of magnitude changes in the mass scale
and power spectrum amplification would not qualitatively change this result.

\begin{figure}[h]
	\centering
	\includegraphics[width=0.92\columnwidth]{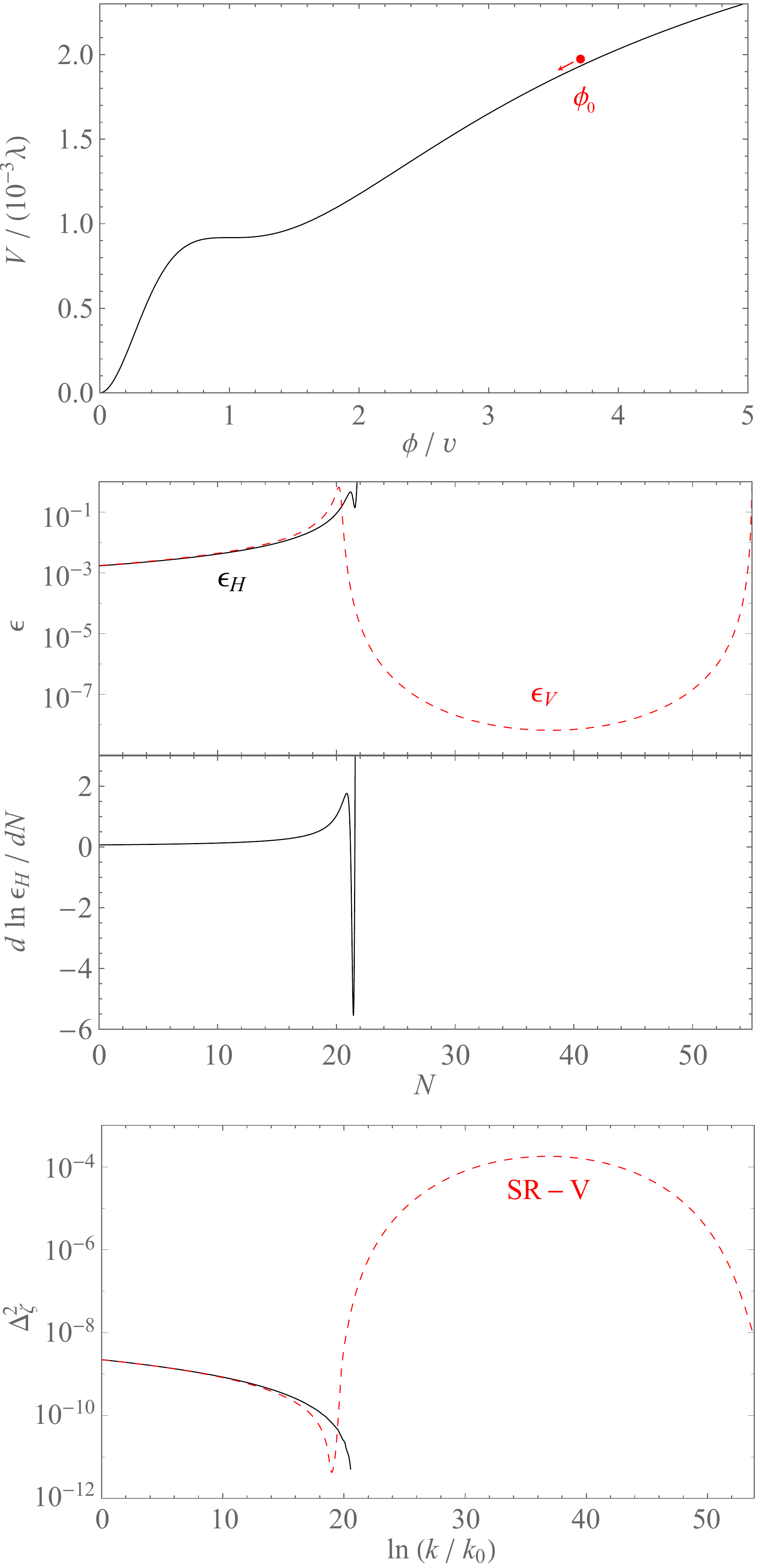}
	\caption{
	Inflection model~\eqref{im-pot} with the parameter set $( a = 3/2, \beta = 4 \times 10^{-5}, \Delta N_{\rm SR} = 35)$~\cite{Garcia-Bellido:2017mdw}.
	Potential (top), evolution of slow-roll parameters (middle), and curvature power spectrum (bottom).  The curvature power spectrum is evaluated by numerical calculation (black, solid) and the potential slow-roll (SR-V) approximation  (red, dashed).  
	}
	\label{fig:im}
\end{figure}

\section{Inflection model}
\label{sec:im}

The inflection model was employed in \cite{Garcia-Bellido:2017mdw}
to attempt to produce a large peak in the curvature power spectrum with a near-inflection point in the potential $V(\phi)$, at which $V'$ and $V''$ are close to zero.  
This means that 
\be \epsilon_V = \frac{1}{2} \left( \frac{V'}{V} \right)^2 \rightarrow 0. \ee
The field equation for the inflaton $\phi$ and the Friedmann equation can be written as
\be \frac{\epsilon_V}{\epsilon_H} =\left( 1 + \frac{1}{2(3-\epsilon_H)} \frac{ d\ln \epsilon_H }{dN}\right)^2
\label{eqn:KG} \ee
and so given the slow-roll condition (\ref{eqn:slowrollviolation})  $\epsilon_V \approx \epsilon_H$.   
To map the potential and its derivative onto the power spectrum in $k$ space, the slow-roll approximation uses 
\be \epsilon_H = \frac{1}{2} \left( \frac{d\phi}{dN} \right)^2 \approx \epsilon_V  \ee
to solve for $N(\phi)$ on the slow-roll attractor 
\be \f{d\phi}{dN} = -\f{V'}{V}.
\label{eqn:phiNSR} \ee
In order to distinguish this solution from the exact relation we call this $N_{\rm SR}(\phi)$.
Since $V \approx 3H^2$, the scalar power spectrum is given by the slow-roll potential form (SR-V) 
\be \Delta_\zeta^2 = \f{V}{24\pi^2 \epsilon_V}\Big|_{k\eta(\phi)=1}, \ee
where $\eta$ is the conformal time to the end of inflation and $\eta \approx k_0^{-1} e^{-N_{\rm SR}(\phi)}$ under SR-V.
An inflection model where $\epsilon_V\rightarrow 0$ would then predict a large
enhancement of the power spectrum under SR-V but violations of (\ref{eqn:slowrollviolation})
can prevent this from occurring in practice.

\begin{figure}[t]
	\centering
	\includegraphics[width=0.92\columnwidth]{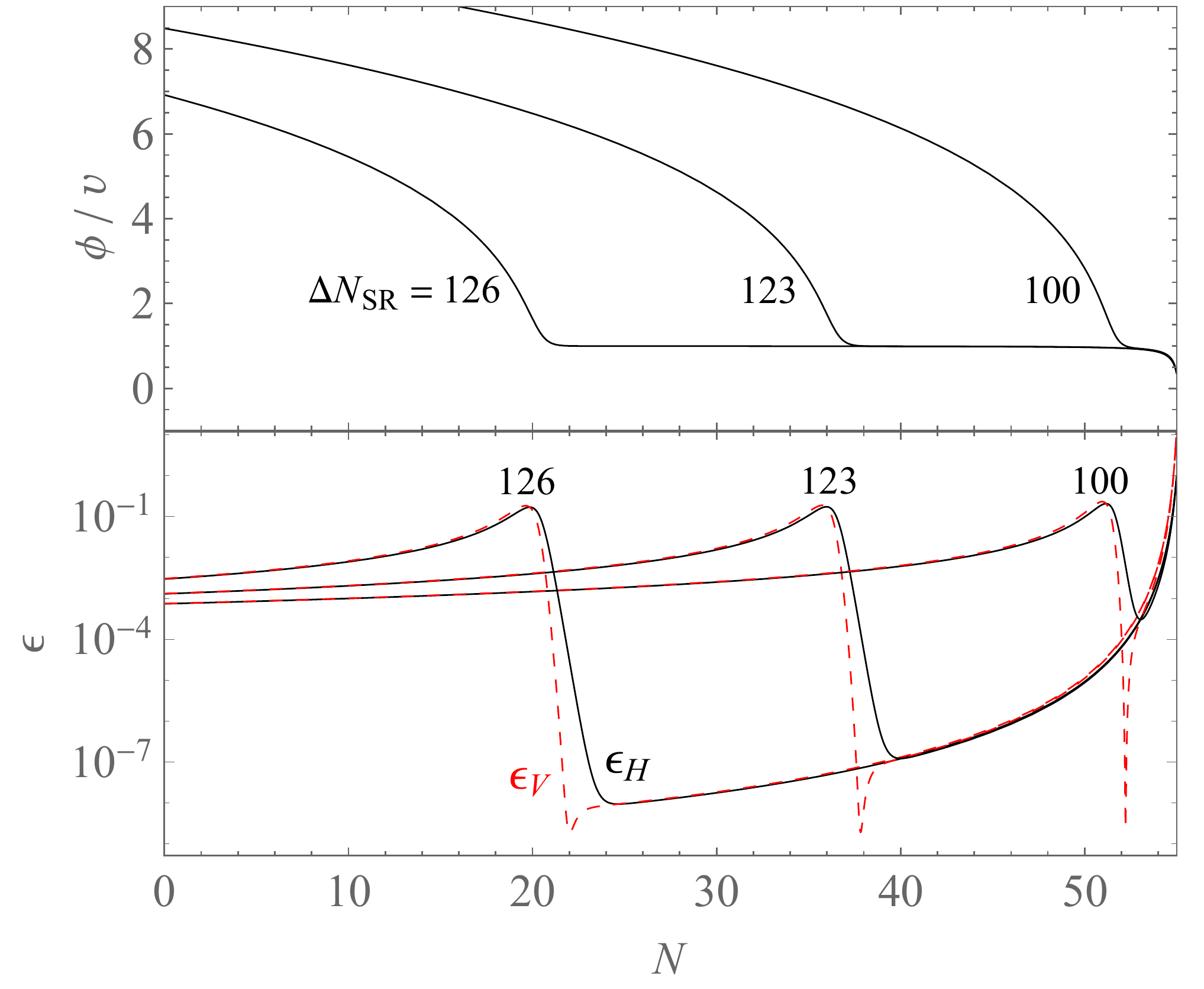}
	\caption{
	Evolution of inflaton and Hubble and potential slow-roll parameters for the inflection model~\eqref{im-pot} with various values of $\Delta N_{\rm SR}$ with the other parameters $( a = 3/2, \beta = 4 \times 10^{-5})$.  While tuning $\Delta N_{\rm SR}$ provides a small window where $\epsilon_H$ is suppressed without the inflaton being stuck near the inflection for too many efolds, all cases fall short of
	the required $10^7$ reduction from the near scale-invariant limit at early times.
	}
	\label{fig:im-tune}
\end{figure}

The inflection potential explored in \cite{Garcia-Bellido:2017mdw} is given by
\be \label{im-pot} V(\phi) = \f{\lambda v^4}{12} \f{x^2 (6 - 4 a x + 3 x^2)}{(1 + b x^2)^2} , \ee
where $x=\phi/v$.  They define  
\begin{align} 
\beta &\equiv b - \kk{ 1 - \f{a^2}{3} + \f{a^2}{3} \mk{\f{9}{2a^2} - 1}^{2/3}  }  , \notag\\
\Delta N_{\rm SR} &\equiv \f{3\pi}{2} \f{v^2}{a^{7/4}} \f{(3/\sqrt{2}-a)^{1/2}}{\sqrt{\beta}} ,
\label{eqn:inflectionparams}
\end{align}
where $\beta=0$ means the existence of an inflection point $V''=0$ at real $x$ and 
$\Delta N_{\rm SR}$ is associated with the efolds required to cross the inflection under the slow-roll
approximation (\ref{eqn:phiNSR}).
The parameter $\lambda$ is determined by the CMB normalization (\ref{CMBnorm}).
Given $\lambda$,  the phenomenological parameters  $(a,\beta,\Delta N_{\rm SR})$ fully define the potential.  
Note that a similar inflection potential is also considered in the context of critical Higgs inflation~\cite{Ezquiaga:2017fvi}.

Following \cite{Garcia-Bellido:2017mdw}, we choose $(a, \beta, \Delta N_{\rm SR} ) = (3/2, 4 \times 10^{-5}, 35)$.
As shown in Fig.~\ref{fig:im}, the potential has a plateau around the near-inflection point $\phi = v$.
Under the slow-roll approximation the inflaton should slow down and take  $\Delta N_{\rm SR} =35$ efolds to cross this point, which would significantly amplify the power spectrum.
The SR-V quantities are shown in Fig.~\ref{fig:im} by red dashed curves.  Here we have assumed that the CMB normalization scale $k_0=0.05\,{\rm Mpc}^{-1}$ exits the horizon at $\phi=\phi_0\approx 3.71$ when $N=0$.

However, in reality the inflaton arrives at the near-inflection point with excess kinetic energy relative to the inflationary attractor  (see Fig.~\ref{fig:im}).  
The maximum violation of slow roll $d\ln \epsilon_H/dN\approx -6$ occurs when
$\epsilon_V \rightarrow 0$ in \eqref{eqn:KG} and the field consequently cannot slow down sufficiently
quickly to stay on the attractor.  Instead the field rolls past this point in about an efold  $\Delta N\approx 1$ and ends inflation 
soon thereafter.
Consequently $\epsilon_H \gg \epsilon_V$ at the near-inflection point and 
the inflation never slows down below $\epsilon_H \sim 0.1$.

Figure~\ref{fig:im} (bottom) compares the power spectra from the SR-V approximation and the exact 
numerical solution of the Mukhanov-Sasaki (MS) equation of motion for the curvature fluctuation.  Whereas the SR-V approximation shows a large enhancement of the power spectrum, 
the true solution does not.    In this case since even the SR approximation of (\ref{SRpower}) correctly
predicts the lack of a power spectrum enhancement and no PBHs, we do not consider the other
approximations of Appendix~\ref{sec:D2fs} further.

While the parameter set $(a, \beta, \Delta N_{\rm SR} ) = (3/2, 4 \times 10^{-5}, 35)$ does not sustain a long enough period for the inflaton to be trapped at the near-inflection point and reduce $\epsilon_H$,
as shown in Fig.~\ref{fig:im-tune}, increasing the parameter $\Delta N_{\rm SR}$ in (\ref{eqn:inflectionparams}) to
$\Delta N_{\rm SR} = 100, 123, 126$ allows the inflaton to traverse the near inflection in $\Delta N \sim 2, 15, 30$, specifically the efolds between entering into the inflection and the end of inflation.   This does lead to $\epsilon_H$ reductions of $2.4, 1.1 \times 10^4, 3.2 \times 10^5$ compared to CMB scales at $N=0$.
There is therefore  a small and fine-tuned window in which the power spectrum is
substantially enhanced without the inflaton getting stuck so that there are too many
efolds between when CMB scales left the horizon and the end of inflation.   Moreover,
the net reduction in $\epsilon_H$ between the nearly scale-invariant region before the inflection to the minimum rapidly saturates at a level that falls short of 
the required $10^7$ enhancement for PBHs to be the dark matter\footnote{
Reference~\cite{Garcia-Bellido:2017mdw} aimed to produce $\sim 10^4$ amplification with the inflection model compared to the $\sim 10^7$ amplification taken here as required for PBHs, which originates from their threshold value $\zeta_c=0.086$.}.

Although further increasing 
$\Delta N_{\rm SR}$ can enhance the change from $N=0$ to the minimum by shifting the curves in Fig.~\ref{fig:im-tune} to the left, it does so
by placing the CMB scales near the maximum of $\epsilon_H$ where the power spectrum has an unacceptably large
red tilt.   We conclude that at least for these values of $a$ and $\beta$, it is impossible to
make PBHs be all of the dark matter in the inflection model.

\begin{figure}[t]
	\centering
	\includegraphics[width=0.92\columnwidth]{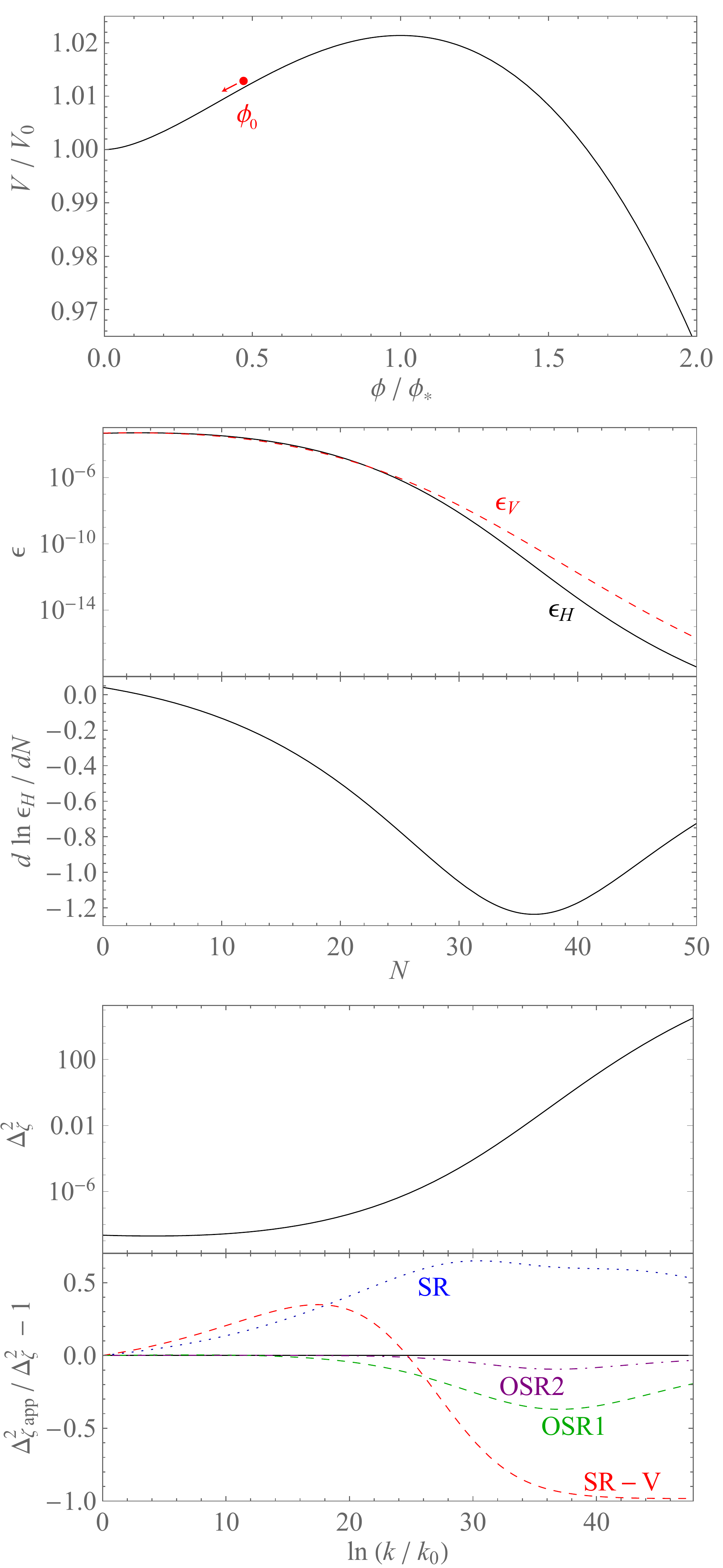}
	\caption{Running mass model (\ref{rm-pot}) with the parameter set \eqref{rm-param}.  Potential (top), evolution of slow-roll parameters (middle), and the curvature power spectrum with the relative error for various approximations (bottom).  
	SR-V remains a poor approximation to the power spectrum enhancement whereas OSR2
	provides accurate results.
	}
	\label{fig:rm}
\end{figure}

\section{Running mass model}
\label{sec:rm}

The logarithmically running mass model provides a more phenomenological approach for designing a potential that produces PBHs.
Let us assume the potential
\be V(\phi) = V_0 + \f{1}{2}m^2(\ln\phi)\phi^2 , \ee
has a local extremum at $\phi=\phi_*$, namely, $V'(\phi_*)=0$.
Using a Taylor expansion of $m^2(\ln\phi)$
around $\ln \phi=\ln \phi_*$, we can express the potential as 
\be \label{rm-pot} V(\phi) = V_0 \kk{ 1 + \f{\tilde c}{8} \phi^2 \mk{ - 1 + 2 L + \tilde g L^2 } },  \ee
where $L\equiv \ln(\phi/\phi_*)$.
Following \cite{Drees:2011hb}, we set  
\be \label{rm-param} \tilde c = -0.1711, \quad 
\tilde g = 0.09648, \quad 
L_0 = -0.756, \ee
where $L_0=\ln(\phi_0/\phi_*)$ and the pivot scale $k_0$ exits the horizon at $\phi=\phi_0\approx 0.47 \phi_*$.
This choice provides a tilt on CMB scales of  $n_s(k_0) = 0.964$ with an
allowed running  $\alpha_s(k_0) = 0.012$ and large running of the running of its value.  This provides
a sufficiently scale-invariant spectrum at CMB scales while allowing a large enhancement of the power spectrum at much higher $k$ for PBH formation.

Since this model is based on Taylor expansion around $\phi_*$, it only applies the vicinity of $\phi_*$ and does not describe how inflation ends.   
Indeed since $\epsilon_H$ is continuously decreasing, if this truncation were exact, the power spectrum would be continuously amplified on small scales and overproduce PBHs.
Hence, we should regard  this model as a phenomenological description of inflation only
between  CMB and the onset of PBH formation where $\Delta_\zeta^2 = {\cal O}(10^{-2})$.

The potential, slow-roll parameters, and power spectrum are presented in Fig.~\ref{fig:rm}.
As before, we normalize efolds as $\phi(N=0)=\phi_0$.
In this model $d\ln \epsilon_H/dN \approx -1.2$ at its extremum,
indicating an ${\cal O}(1)$  violation of (\ref{eqn:slowrollviolation}). 
Consequently, even though at the same field position $\epsilon_V(\phi) \approx \epsilon_H(N(\phi))$, the prolonged violation causes the slow-roll approximation 
(\ref{eqn:phiNSR}) to misestimate the efold corresponding to a given field position
$N_{\rm SR}(\phi) \ne N(\phi)$.   In the middle panel of Fig.~\ref{fig:rm}, we see that this leads to a strong deviation of $\epsilon_V(N_{\rm SR})$ from $\epsilon_H(N)$.

The bottom panel of Fig.~\ref{fig:rm} depicts the numerical solution for the curvature power spectrum
and the relative error for the various formulae given in Appendix~\ref{sec:D2fs}. 
The curvature power spectrum is enhanced to a sufficient value to form PBHs 
$\Delta_\zeta^2 =0.02$ by $\ln(k/k_0)\approx35$.
We see that SR-V underestimates the power spectrum by a factor of $8.4$ ($-88$\%)
there and substantially more beyond this point.

The SR approximation of (\ref{SRpower}) corrects $\phi(N)$ and $\epsilon_H$ and performs better with a maximal overestimate  of a factor of 1.7 (70\%) at $N\approx 30$.  Most of this improvement comes from
the correction to $\phi(N)$ and note that the SR-V case integrates (\ref{eqn:phiNSR}), an approach that is called ``exact" in  \cite{Drees:2011hb}.  While this misestimate amounts mainly
to a shift in efolds, the efolds from the CMB scale control the mass scale of
the PBHs through (\ref{Nm}).
On the other hand, the errors for the optimized first and second order Hubble slow roll approximations
of Appendix~\ref{sec:D2fs}, OSR1 and OSR2, peak at about $40\%$ and $10\%$, respectively.
Since $d\ln\epsilon_H/dN$ never greatly exceeds unity, the higher approximations improve the accuracy without adding much to the computational cost.

\begin{figure}[t]
	\centering
	\includegraphics[width=0.92\columnwidth]{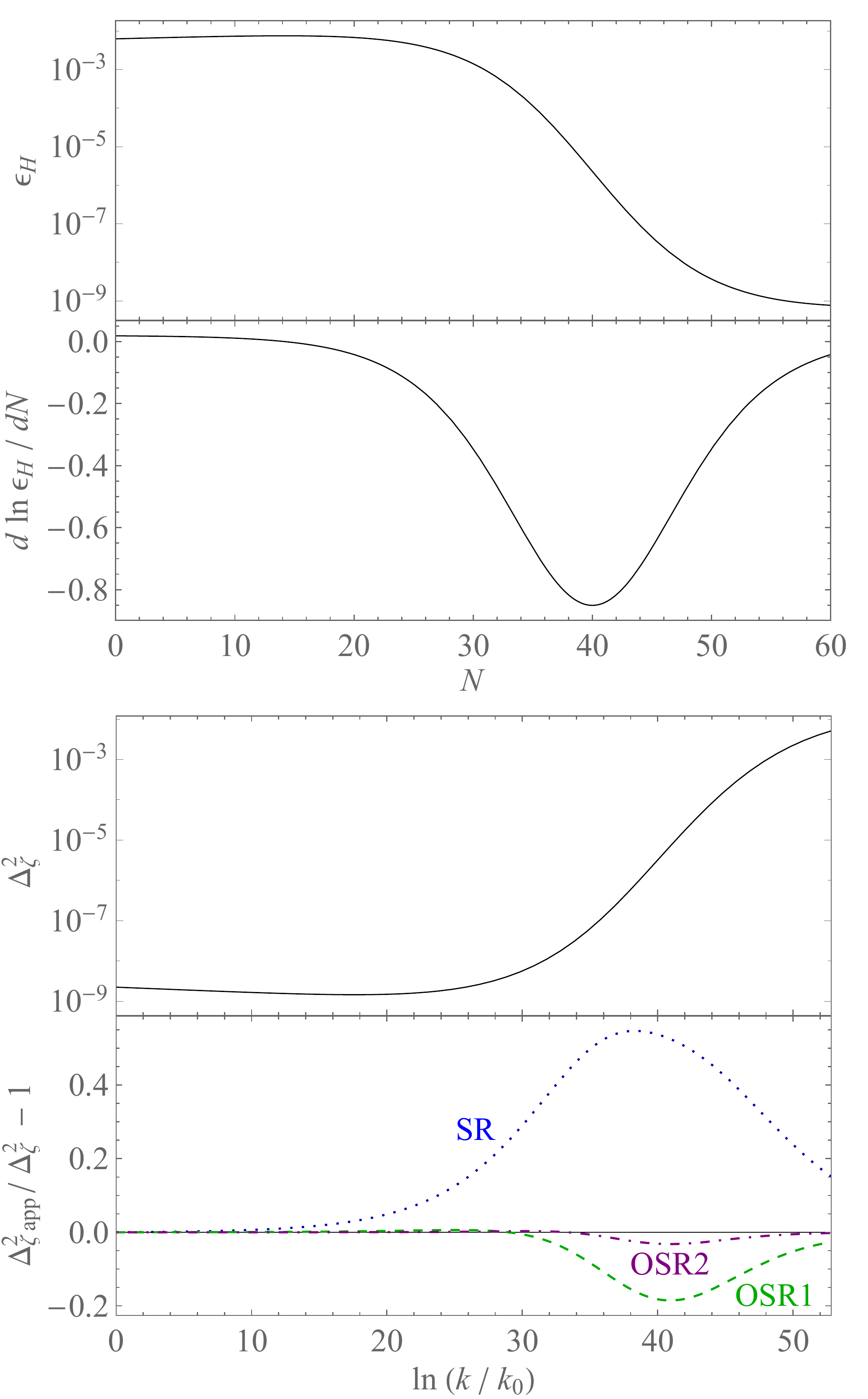}
	\caption{Step model  \eqref{tanh} with the parameter set $(\Ca,\Cb, \Cc,\Ns,d)=(-5.07,0.0194,8.7,40,10)$.
	Evolution of slow-roll parameter (top) and curvature power spectrum with the relative error for various approximations (bottom).  The relatively wide width $d$ produces slow-roll violation that
	can be described to better than $\sim 4\%$ accuracy by the  OSR2 approximation.
	}
	\label{fig:tanhd10}
\end{figure}

\section{Slow-roll Step Model}
\label{sec:eps}

In order to more systematically  explore slow-roll violation in the PBH context, we can bypass the choice of a potential and directly model the critical function $\epsilon_H(N)$, as in an effective field theory of inflation approach.  This approach also allows one to straightforwardly generalize these results to 
noncanonical inflation models in the Horndeski and Gleyzes-Langlois-Piazza-Vernizzi (GLPV) classes by replacing $\epsilon_H(N)$
with the general scalar source function~\cite{Motohashi:2017gqb}.

CMB measurements place observational constraints on $\epsilon_H$ around $N=0$, the
epoch when $k_0=0.05$ Mpc$^{-1}$ left the horizon during inflation.  If the slow-roll 
approximation is satisfied on these scales $\epsilon_H$ should be well characterized locally by its Taylor expansion
\be \ln \epsilon_H \approx \Ca + \Cb N , \quad (N\approx 0). \ee
The scalar tilt  and the tensor-scalar ratio are given by
\begin{align} \label{epsHparams}
n_s-1&\approx -2\epsilon_H -\frac{d\ln \epsilon_H}{dN}
= -2e^{\Ca} -\Cb ,\notag\\
r&\approx 16\epsilon_H
=16e^{\Ca} .
\end{align}
The Planck CMB data imply \cite{Ade:2015lrj}
\be n_s\approx 0.968, \quad r<0.10 . \ee
For definiteness, let us take the upper bound on $r$ to fix $\Ca$, and then use $n_s$ to fix $\Cb$:
\be \Ca = -5.07,\quad \Cb=0.0194.  \ee

Next, for PBH formation we must change $\ln\epsilon_H$ by at least $\ln( 10^7 )\sim 16$ 
before the end of inflation.   We therefore consider a steplike transition around $\Ns$
\be \label{tanh} \ln \epsilon_H = \Ca + \Cb N - \Cc \kk{ 1 + \tanh \mk{\f{N-\Ns}{d}} } , \ee
Similar to the running mass model, we assume that this form only parametrizes
$N \lesssim \Ns$ so that $\ln \epsilon_H$ undergoes another transition to end inflation near $N=60$.
The step causes a change of $\Delta\ln \epsilon_H \sim -2\Cc$ across $\sim d$ efolds.
For definiteness, we take  $\Cc= 8.7$ and $\Ns=40$ so that $d$ determines the amount by which
(\ref{eqn:slowrollviolation}) is violated.

In Fig.~\ref{fig:tanhd10}, we first consider a fairly wide step 
\be (\Ca,\Cb, \Cc,\Ns,d)=(-5.07,0.0194,8.7,40,10). \ee  
Here $\ln \epsilon_H$ has its minimum at $N\approx 65.9$, and its difference from $N=0$ is $\Delta \ln \epsilon_H \approx 16.0$.
The maximum amplitude of slow-roll violation is $d\ln\epsilon_H/dN\approx -0.85$.
The relative errors for SR, OSR1, OSR2 are about $60\%$, $-20\%$, $-4\%$, respectively.
For slow changes of $\ln \epsilon_H$, the optimized formula works extremely well.

\begin{figure}[t]
	\centering
	\includegraphics[width=0.92\columnwidth]{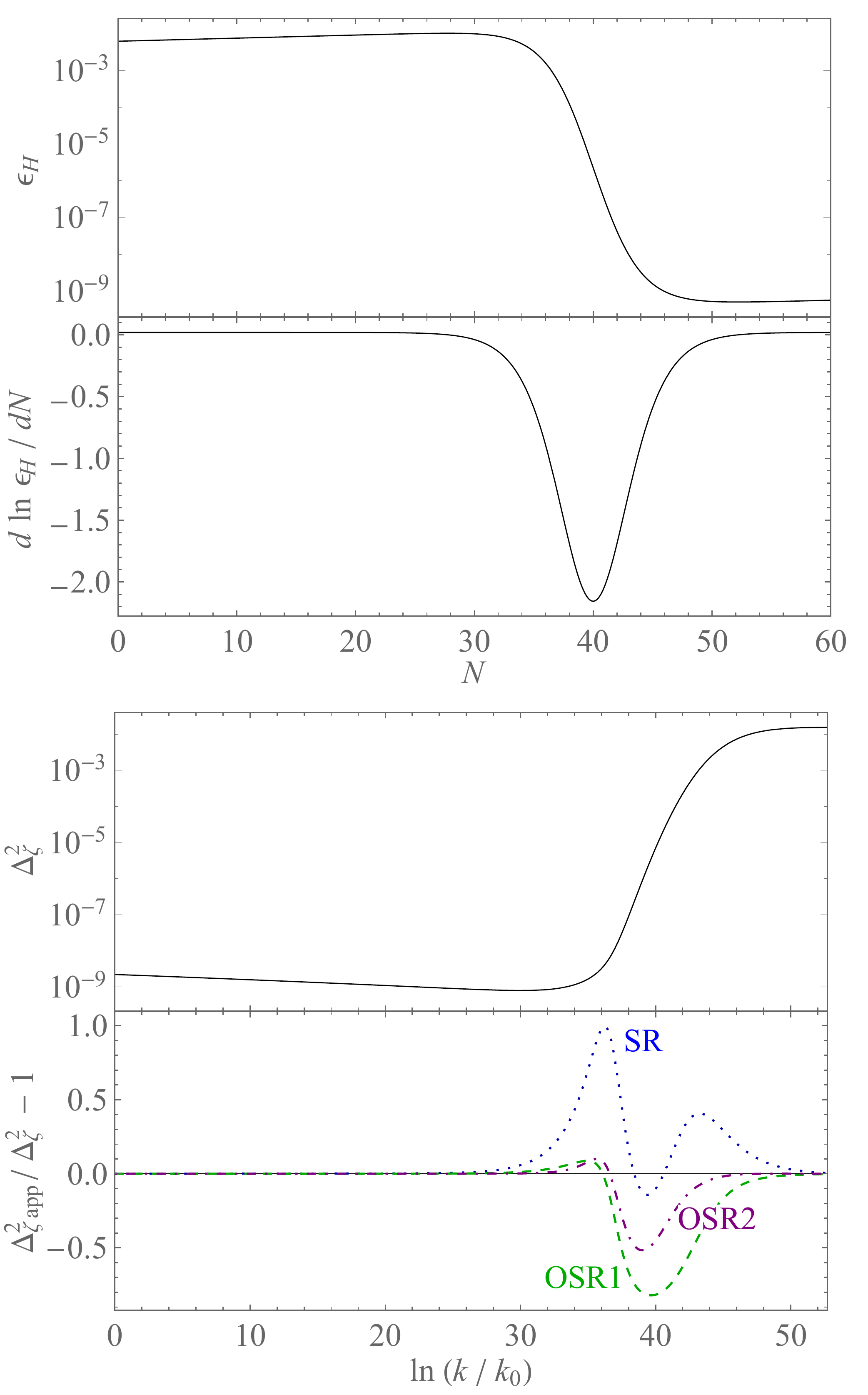}
	\caption{Step model  \eqref{tanh} with the parameter set $(\Ca,\Cb, \Cc,\Ns,d)=(-5.07,0.0194,8.7,40,4)$.
	Evolution of slow-roll parameter (top) and curvature power spectrum with the relative error for various approximations (bottom). As the width $d$ decreases, the slow-roll violations increase, reducing the efficacy of higher order corrections.
	}
	\label{fig:tanhd4}
\end{figure}

We next increase the slow-roll violation by decreasing the step width in Fig.~\ref{fig:tanhd4},  
\be (\Ca,\Cb,\Cc,\Ns,d)=(-5.07,0.0194,8.7,40,4), \ee  
for which $\ln \epsilon_H$ has its minimum at $N\approx 52.2$, and its difference from $N=0$ is  $\Delta \ln \epsilon_H \approx 16.3$.
The maximum amplitude of time variation is $d\ln\epsilon_H/dN\approx -2.2$.
The relative errors for SR, OSR1, OSR2 are about $100\%$, $-80\%$, $-50\%$, respectively.  
While OSR2 still performs relatively well, the larger the violation of slow roll, the less the higher order optimizations improve the result.
For the same level of slow-roll violation $\Cc/d$, 
the approximations perform similarly if the amplitude $\Cc$ is changed at fixed width $d$.

\section{Conclusion}
\label{sec:conc}

For PBHs to be the dark matter in single-field inflation, the slow-roll approximation must be violated by at least ${\cal O}(1)$
in order to enhance the curvature power spectrum within the required number of efolds 
between CMB  and PBH  scales.
As a consequence, power spectrum predictions which rely on the inflaton remaining
on the slow-roll attractor can fail dramatically.   
Models like the inflection potential which might seem to enhance the power spectrum
under the potential slow-roll approximation when calculated properly provide no enhancement
across most of its parameter space.

Since the slow-roll approximation must fail by ${\cal O}(1)$ but is not required to fail by
much more, approximations based on an optimized temporal evaluation of Hubble slow-roll
parameters at various orders (OSR1,2) can perform much better at little extra
computational cost.   In particular the OSR2 approximation remains a good description 
of the potential out to slow-roll violations of  $|d\ln\epsilon_H/dN |<2$ which encompasses a wide range of PBH
formation models where the $10^7$ amplification of power occurs in $\sim 10$ efolds or more.

\vspace{5mm}

\noindent
{\bf Note added:} 
The analysis of the inflection model in \S\ref{sec:im} is based on the parameter set used in  version 3 of \cite{Garcia-Bellido:2017mdw}.   In  version 4 of \cite{Garcia-Bellido:2017mdw}, which appeared 
after this work was completed, a  different set of parameters is used instead  to produce a large amplification, for which the difference between SR-V and exact calculation of curvature power spectrum is reduced. 
As mentioned in the last two paragraphs of \S\ref{sec:im}, this type of modification requires tuning and still fails to
produce the required $10^7$ amplification of power over the nearly scale-invariant portion of the
curvature power spectrum.

While this work was being completed, \cite{Kannike:2017bxn} appeared where the violation of slow roll in inflection model was mentioned in Sec.~2.3.  Also, \cite{Germani:2017bcs} investigated the violation of slow-roll approximation and an amplification of the curvature power spectrum during ultra-slow-roll phase in the inflection models in~\cite{Garcia-Bellido:2017mdw,Ezquiaga:2017fvi} reaching similar conclusions.

\acknowledgments
H.M.\ was supported in part by MINECO Grant SEV-2014-0398,
PROMETEO II/2014/050,
Spanish Grant FPA2014-57816-P of the MINECO, and
European Union’s Horizon 2020 research and innovation programme under the Marie Sk\l{}odowska-Curie grant agreements No.~690575 and 674896.
W.H.\ was supported by grants 
NSF PHY-0114422, 
NSF PHY-0551142,
U.S.~Dept.\ of Energy contract DE-FG02-13ER41958, and 
NASA ATP NNX15AK22G.

\appendix

\section{Power Spectrum Approximations}
\label{sec:D2fs}

In this appendix, we summarize various power spectrum 
approximations compared in the main text.
For the derivation of OSR, see \cite{Motohashi:2015hpa} for details.

The Hubble flow slow-roll parameters are defined by
\begin{align}
\epsilon_H&\equiv - \f{d\ln H}{dN}, \notag\\
\delta_1&\equiv \f{1}{2} \f{d\ln \epsilon_H}{dN} - \epsilon_H ,\notag\\
\delta_2 &\equiv \f{d\delta_1}{dN} + \delta_1(\delta_1-\epsilon_H) ,
\end{align}
whereas the potential slow-roll parameter is
\be \epsilon_V\equiv \f{1}{2}\mk{\f{V'}{V}}^2 . \ee

\begin{itemize}[leftmargin=0.2in]

\item SR-V
\be \label{eq:SR0-V} \Delta_\zeta^2 = \left. \f{V}{24\pi^2 \epsilon_V} \right|_{k\eta=1} , \ee
with background evolution $\phi(N)$ solved on the slow-roll attractor $\epsilon_V\approx\epsilon_H$
\be \f{d\phi}{dN} = - \f{V'}{V} , \ee
and $\eta(N)$ given by the Friedmann equation.  In this approximation, the power spectrum can be simply estimated from
the potential and $\phi(N)$ since $3 H^2 \approx V \approx$\,const.\ so that $k\eta \approx e^{-N(\phi)} k/k_0$.   

\item SR
\be \label{eq:SR0-H} \Delta_\zeta^2 = \left. \f{H^2}{8\pi^2 \epsilon_H} \right|_{k\eta=1} \ee
The standard leading-order slow-roll approximation.  
It can be evaluated by solving background evolution equations.



\item OSR1
\be \ln \Delta_\zeta^2 = \left. \ln \f{H^2}{8\pi^2 \epsilon_H} - \f{10}{3}\epsilon_H - \f{2}{3} \delta_1 \right|_{k\eta=x_1} . \ee
The leading order optimized slow-roll formula obtained from Taylor expansion of the generalized slow-roll approximation~\cite{Motohashi:2015hpa}.  
The evaluation epoch is not horizon exit $k\eta=1$ but $k\eta=x_1$ where
\be \ln x_1 \equiv \f{7}{3} - \ln 2 -\gamma_E, \ee
which corresponds to resumming the next-leading-order term in the source evolution.
In \cite{Motohashi:2015hpa} it was denoted as the optimized leading-order approximation (OLO).

\item OSR2
\begin{align} 
& \ln \Delta_\zeta^2
= \ln \f{H^2}{8\pi^2\epsilon_H} - \f{10}{3} \epsilon_H - \f{2}{3} \delta_1 - \f{23}{3} \epsilon_H^2 - 6\epsilon_H \delta_1 \notag\\
&~~~ + q_1 \mk{ 4\epsilon_H + 2\delta_1 + \f{2}{3} \delta_2 + \f{32}{3} \epsilon_H^2 + \f{28}{3} \epsilon_H \delta_1 - \f{2}{3} \delta_1^2 } \notag\\
&~~~ + \left. \mk{ \f{\pi^2}{2} -4 } (2\epsilon_H + \delta_1)^2 \right|_{k\eta=x_2} ,
\end{align}
where
\begin{align} 
\ln x_2 &\equiv \f{7}{3} - \ln 2 -\gamma_E - \f{ \sqrt{3 \pi^2 - 4} }{6} ,\notag\\
q_1(\ln x_2) &= \f{ \sqrt{3 \pi^2 - 4} }{6}.
\end{align}
The next-leading-order optimized slow-roll formula obtained from the Taylor expansion of
the generalized slow-roll approximation.  
In \cite{Motohashi:2015hpa} it was denoted as the optimized next-to-leading-order approximation (ONO).

\item Numerical (MS)
\be \Delta_\zeta^2 = \left. \f{k^3}{2\pi^2}|\zeta_k|^2 \right|_{k\eta \ll 1} . \ee
Numerical solution of Mukhanov-Sasaki equation of the curvature modefunction $\zeta_k$ with Bunch-Davies boundary condition at $k\eta \gg 1$.

\end{itemize}

\vskip 5cm

\bibliography{ref-PBH}

\begin{thebibliography}{35}%
\makeatletter
\providecommand \@ifxundefined [1]{%
 \@ifx{#1\undefined}
}%
\providecommand \@ifnum [1]{%
 \ifnum #1\expandafter \@firstoftwo
 \else \expandafter \@secondoftwo
 \fi
}%
\providecommand \@ifx [1]{%
 \ifx #1\expandafter \@firstoftwo
 \else \expandafter \@secondoftwo
 \fi
}%
\providecommand \natexlab [1]{#1}%
\providecommand \enquote  [1]{``#1''}%
\providecommand \bibnamefont  [1]{#1}%
\providecommand \bibfnamefont [1]{#1}%
\providecommand \citenamefont [1]{#1}%
\providecommand \href@noop [0]{\@secondoftwo}%
\providecommand \href [0]{\begingroup \@sanitize@url \@href}%
\providecommand \@href[1]{\@@startlink{#1}\@@href}%
\providecommand \@@href[1]{\endgroup#1\@@endlink}%
\providecommand \@sanitize@url [0]{\catcode `\\12\catcode `\$12\catcode
  `\&12\catcode `\#12\catcode `\^12\catcode `\_12\catcode `\%12\relax}%
\providecommand \@@startlink[1]{}%
\providecommand \@@endlink[0]{}%
\providecommand \url  [0]{\begingroup\@sanitize@url \@url }%
\providecommand \@url [1]{\endgroup\@href {#1}{\urlprefix }}%
\providecommand \urlprefix  [0]{URL }%
\providecommand \Eprint [0]{\href }%
\providecommand \doibase [0]{http://dx.doi.org/}%
\providecommand \selectlanguage [0]{\@gobble}%
\providecommand \bibinfo  [0]{\@secondoftwo}%
\providecommand \bibfield  [0]{\@secondoftwo}%
\providecommand \translation [1]{[#1]}%
\providecommand \BibitemOpen [0]{}%
\providecommand \bibitemStop [0]{}%
\providecommand \bibitemNoStop [0]{.\EOS\space}%
\providecommand \EOS [0]{\spacefactor3000\relax}%
\providecommand \BibitemShut  [1]{\csname bibitem#1\endcsname}%
\let\auto@bib@innerbib\@empty
\bibitem [{\citenamefont {Abbott}\ \emph
  {et~al.}(2016{\natexlab{a}})\citenamefont {Abbott} \emph
  {et~al.}}]{Abbott:2016blz}%
  \BibitemOpen
  \bibfield  {author} {\bibinfo {author} {\bibfnamefont {B.~P.}\ \bibnamefont
  {Abbott}} \emph {et~al.} (\bibinfo {collaboration} {Virgo, LIGO
  Scientific}),\ }\href {\doibase 10.1103/PhysRevLett.116.061102} {\bibfield
  {journal} {\bibinfo  {journal} {Phys. Rev. Lett.}\ }\textbf {\bibinfo
  {volume} {116}},\ \bibinfo {pages} {061102} (\bibinfo {year}
  {2016}{\natexlab{a}})},\ \Eprint {http://arxiv.org/abs/1602.03837}
  {arXiv:1602.03837 [gr-qc]} \BibitemShut {NoStop}%
\bibitem [{\citenamefont {Abbott}\ \emph
  {et~al.}(2016{\natexlab{b}})\citenamefont {Abbott} \emph
  {et~al.}}]{Abbott:2016nmj}%
  \BibitemOpen
  \bibfield  {author} {\bibinfo {author} {\bibfnamefont {B.~P.}\ \bibnamefont
  {Abbott}} \emph {et~al.} (\bibinfo {collaboration} {Virgo, LIGO
  Scientific}),\ }\href {\doibase 10.1103/PhysRevLett.116.241103} {\bibfield
  {journal} {\bibinfo  {journal} {Phys. Rev. Lett.}\ }\textbf {\bibinfo
  {volume} {116}},\ \bibinfo {pages} {241103} (\bibinfo {year}
  {2016}{\natexlab{b}})},\ \Eprint {http://arxiv.org/abs/1606.04855}
  {arXiv:1606.04855 [gr-qc]} \BibitemShut {NoStop}%
\bibitem [{\citenamefont {Abbott}\ \emph {et~al.}(2017)\citenamefont {Abbott}
  \emph {et~al.}}]{Abbott:2017vtc}%
  \BibitemOpen
  \bibfield  {author} {\bibinfo {author} {\bibfnamefont {B.~P.}\ \bibnamefont
  {Abbott}} \emph {et~al.} (\bibinfo {collaboration} {VIRGO, LIGO
  Scientific}),\ }\href {\doibase 10.1103/PhysRevLett.118.221101} {\bibfield
  {journal} {\bibinfo  {journal} {Phys. Rev. Lett.}\ }\textbf {\bibinfo
  {volume} {118}},\ \bibinfo {pages} {221101} (\bibinfo {year} {2017})},\
  \Eprint {http://arxiv.org/abs/1706.01812} {arXiv:1706.01812 [gr-qc]}
  \BibitemShut {NoStop}%
\bibitem [{\citenamefont {{Zel'dovich}}\ and\ \citenamefont
  {{Novikov}}(1966)}]{1966AZh....43..758Z}%
  \BibitemOpen
  \bibfield  {author} {\bibinfo {author} {\bibfnamefont {Y.~B.}\ \bibnamefont
  {{Zel'dovich}}}\ and\ \bibinfo {author} {\bibfnamefont {I.~D.}\ \bibnamefont
  {{Novikov}}},\ }\href {http://adsabs.harvard.edu/abs/1966AZh....43..758Z}
  {\bibfield  {journal} {\bibinfo  {journal} {Astron. Zh.}\ }\textbf {\bibinfo
  {volume} {43}},\ \bibinfo {pages} {758} (\bibinfo {year} {1966})}\BibitemShut
  {NoStop}%
\bibitem [{\citenamefont {Hawking}(1971)}]{Hawking:1971ei}%
  \BibitemOpen
  \bibfield  {author} {\bibinfo {author} {\bibfnamefont {S.}~\bibnamefont
  {Hawking}},\ }\href {\doibase 10.1093/mnras/152.1.75} {\bibfield  {journal}
  {\bibinfo  {journal} {Mon. Not. Roy. Astron. Soc.}\ }\textbf {\bibinfo
  {volume} {152}},\ \bibinfo {pages} {75} (\bibinfo {year} {1971})}\BibitemShut
  {NoStop}%
\bibitem [{\citenamefont {Carr}(1975)}]{Carr:1975qj}%
  \BibitemOpen
  \bibfield  {author} {\bibinfo {author} {\bibfnamefont {B.~J.}\ \bibnamefont
  {Carr}},\ }\href {\doibase 10.1086/153853} {\bibfield  {journal} {\bibinfo
  {journal} {Astrophys. J.}\ }\textbf {\bibinfo {volume} {201}},\ \bibinfo
  {pages} {1} (\bibinfo {year} {1975})}\BibitemShut {NoStop}%
\bibitem [{\citenamefont {Bird}\ \emph {et~al.}(2016)\citenamefont {Bird},
  \citenamefont {Cholis}, \citenamefont {Muñoz}, \citenamefont {Ali-Haïmoud},
  \citenamefont {Kamionkowski}, \citenamefont {Kovetz}, \citenamefont
  {Raccanelli},\ and\ \citenamefont {Riess}}]{Bird:2016dcv}%
  \BibitemOpen
  \bibfield  {author} {\bibinfo {author} {\bibfnamefont {S.}~\bibnamefont
  {Bird}}, \bibinfo {author} {\bibfnamefont {I.}~\bibnamefont {Cholis}},
  \bibinfo {author} {\bibfnamefont {J.~B.}\ \bibnamefont {Muñoz}}, \bibinfo
  {author} {\bibfnamefont {Y.}~\bibnamefont {Ali-Haïmoud}}, \bibinfo {author}
  {\bibfnamefont {M.}~\bibnamefont {Kamionkowski}}, \bibinfo {author}
  {\bibfnamefont {E.~D.}\ \bibnamefont {Kovetz}}, \bibinfo {author}
  {\bibfnamefont {A.}~\bibnamefont {Raccanelli}}, \ and\ \bibinfo {author}
  {\bibfnamefont {A.~G.}\ \bibnamefont {Riess}},\ }\href {\doibase
  10.1103/PhysRevLett.116.201301} {\bibfield  {journal} {\bibinfo  {journal}
  {Phys. Rev. Lett.}\ }\textbf {\bibinfo {volume} {116}},\ \bibinfo {pages}
  {201301} (\bibinfo {year} {2016})},\ \Eprint
  {http://arxiv.org/abs/1603.00464} {arXiv:1603.00464 [astro-ph.CO]}
  \BibitemShut {NoStop}%
\bibitem [{\citenamefont {Clesse}\ and\ \citenamefont
  {García-Bellido}(2017)}]{Clesse:2016vqa}%
  \BibitemOpen
  \bibfield  {author} {\bibinfo {author} {\bibfnamefont {S.}~\bibnamefont
  {Clesse}}\ and\ \bibinfo {author} {\bibfnamefont {J.}~\bibnamefont
  {García-Bellido}},\ }\href {\doibase 10.1016/j.dark.2016.10.002} {\bibfield
  {journal} {\bibinfo  {journal} {Phys. Dark Univ.}\ }\textbf {\bibinfo
  {volume} {15}},\ \bibinfo {pages} {142} (\bibinfo {year} {2017})},\ \Eprint
  {http://arxiv.org/abs/1603.05234} {arXiv:1603.05234 [astro-ph.CO]}
  \BibitemShut {NoStop}%
\bibitem [{\citenamefont {Sasaki}\ \emph {et~al.}(2016)\citenamefont {Sasaki},
  \citenamefont {Suyama}, \citenamefont {Tanaka},\ and\ \citenamefont
  {Yokoyama}}]{Sasaki:2016jop}%
  \BibitemOpen
  \bibfield  {author} {\bibinfo {author} {\bibfnamefont {M.}~\bibnamefont
  {Sasaki}}, \bibinfo {author} {\bibfnamefont {T.}~\bibnamefont {Suyama}},
  \bibinfo {author} {\bibfnamefont {T.}~\bibnamefont {Tanaka}}, \ and\ \bibinfo
  {author} {\bibfnamefont {S.}~\bibnamefont {Yokoyama}},\ }\href {\doibase
  10.1103/PhysRevLett.117.061101} {\bibfield  {journal} {\bibinfo  {journal}
  {Phys. Rev. Lett.}\ }\textbf {\bibinfo {volume} {117}},\ \bibinfo {pages}
  {061101} (\bibinfo {year} {2016})},\ \Eprint
  {http://arxiv.org/abs/1603.08338} {arXiv:1603.08338 [astro-ph.CO]}
  \BibitemShut {NoStop}%
\bibitem [{\citenamefont {Kawasaki}\ \emph {et~al.}(2016)\citenamefont
  {Kawasaki}, \citenamefont {Kusenko}, \citenamefont {Tada},\ and\
  \citenamefont {Yanagida}}]{Kawasaki:2016pql}%
  \BibitemOpen
  \bibfield  {author} {\bibinfo {author} {\bibfnamefont {M.}~\bibnamefont
  {Kawasaki}}, \bibinfo {author} {\bibfnamefont {A.}~\bibnamefont {Kusenko}},
  \bibinfo {author} {\bibfnamefont {Y.}~\bibnamefont {Tada}}, \ and\ \bibinfo
  {author} {\bibfnamefont {T.~T.}\ \bibnamefont {Yanagida}},\ }\href {\doibase
  10.1103/PhysRevD.94.083523} {\bibfield  {journal} {\bibinfo  {journal} {Phys.
  Rev.}\ }\textbf {\bibinfo {volume} {D94}},\ \bibinfo {pages} {083523}
  (\bibinfo {year} {2016})},\ \Eprint {http://arxiv.org/abs/1606.07631}
  {arXiv:1606.07631 [astro-ph.CO]} \BibitemShut {NoStop}%
\bibitem [{\citenamefont {Carr}\ \emph {et~al.}(2016)\citenamefont {Carr},
  \citenamefont {Kuhnel},\ and\ \citenamefont {Sandstad}}]{Carr:2016drx}%
  \BibitemOpen
  \bibfield  {author} {\bibinfo {author} {\bibfnamefont {B.}~\bibnamefont
  {Carr}}, \bibinfo {author} {\bibfnamefont {F.}~\bibnamefont {Kuhnel}}, \ and\
  \bibinfo {author} {\bibfnamefont {M.}~\bibnamefont {Sandstad}},\ }\href
  {\doibase 10.1103/PhysRevD.94.083504} {\bibfield  {journal} {\bibinfo
  {journal} {Phys. Rev.}\ }\textbf {\bibinfo {volume} {D94}},\ \bibinfo {pages}
  {083504} (\bibinfo {year} {2016})},\ \Eprint
  {http://arxiv.org/abs/1607.06077} {arXiv:1607.06077 [astro-ph.CO]}
  \BibitemShut {NoStop}%
\bibitem [{\citenamefont {Inomata}\ \emph {et~al.}(2017)\citenamefont
  {Inomata}, \citenamefont {Kawasaki}, \citenamefont {Mukaida}, \citenamefont
  {Tada},\ and\ \citenamefont {Yanagida}}]{Inomata:2016rbd}%
  \BibitemOpen
  \bibfield  {author} {\bibinfo {author} {\bibfnamefont {K.}~\bibnamefont
  {Inomata}}, \bibinfo {author} {\bibfnamefont {M.}~\bibnamefont {Kawasaki}},
  \bibinfo {author} {\bibfnamefont {K.}~\bibnamefont {Mukaida}}, \bibinfo
  {author} {\bibfnamefont {Y.}~\bibnamefont {Tada}}, \ and\ \bibinfo {author}
  {\bibfnamefont {T.~T.}\ \bibnamefont {Yanagida}},\ }\href {\doibase
  10.1103/PhysRevD.95.123510} {\bibfield  {journal} {\bibinfo  {journal} {Phys.
  Rev.}\ }\textbf {\bibinfo {volume} {D95}},\ \bibinfo {pages} {123510}
  (\bibinfo {year} {2017})},\ \Eprint {http://arxiv.org/abs/1611.06130}
  {arXiv:1611.06130 [astro-ph.CO]} \BibitemShut {NoStop}%
\bibitem [{\citenamefont {Nakama}\ \emph {et~al.}(2017)\citenamefont {Nakama},
  \citenamefont {Silk},\ and\ \citenamefont {Kamionkowski}}]{Nakama:2016gzw}%
  \BibitemOpen
  \bibfield  {author} {\bibinfo {author} {\bibfnamefont {T.}~\bibnamefont
  {Nakama}}, \bibinfo {author} {\bibfnamefont {J.}~\bibnamefont {Silk}}, \ and\
  \bibinfo {author} {\bibfnamefont {M.}~\bibnamefont {Kamionkowski}},\ }\href
  {\doibase 10.1103/PhysRevD.95.043511} {\bibfield  {journal} {\bibinfo
  {journal} {Phys. Rev.}\ }\textbf {\bibinfo {volume} {D95}},\ \bibinfo {pages}
  {043511} (\bibinfo {year} {2017})},\ \Eprint
  {http://arxiv.org/abs/1612.06264} {arXiv:1612.06264 [astro-ph.CO]}
  \BibitemShut {NoStop}%
\bibitem [{\citenamefont {Kühnel}\ and\ \citenamefont
  {Freese}(2017)}]{Kuhnel:2017pwq}%
  \BibitemOpen
  \bibfield  {author} {\bibinfo {author} {\bibfnamefont {F.}~\bibnamefont
  {Kühnel}}\ and\ \bibinfo {author} {\bibfnamefont {K.}~\bibnamefont
  {Freese}},\ }\href {\doibase 10.1103/PhysRevD.95.083508} {\bibfield
  {journal} {\bibinfo  {journal} {Phys. Rev.}\ }\textbf {\bibinfo {volume}
  {D95}},\ \bibinfo {pages} {083508} (\bibinfo {year} {2017})},\ \Eprint
  {http://arxiv.org/abs/1701.07223} {arXiv:1701.07223 [astro-ph.CO]}
  \BibitemShut {NoStop}%
\bibitem [{\citenamefont {Chiba}\ and\ \citenamefont
  {Yokoyama}(2017)}]{Chiba:2017rvs}%
  \BibitemOpen
  \bibfield  {author} {\bibinfo {author} {\bibfnamefont {T.}~\bibnamefont
  {Chiba}}\ and\ \bibinfo {author} {\bibfnamefont {S.}~\bibnamefont
  {Yokoyama}},\ }\href {\doibase 10.1093/ptep/ptx087} {\bibfield  {journal}
  {\bibinfo  {journal} {PTEP}\ }\textbf {\bibinfo {volume} {8}},\ \bibinfo
  {pages} {083} (\bibinfo {year} {2017})},\ \Eprint
  {http://arxiv.org/abs/1704.06573} {arXiv:1704.06573 [gr-qc]} \BibitemShut
  {NoStop}%
\bibitem [{\citenamefont {Carr}\ \emph {et~al.}(2017)\citenamefont {Carr},
  \citenamefont {Raidal}, \citenamefont {Tenkanen}, \citenamefont {Vaskonen},\
  and\ \citenamefont {Veermäe}}]{Carr:2017jsz}%
  \BibitemOpen
  \bibfield  {author} {\bibinfo {author} {\bibfnamefont {B.}~\bibnamefont
  {Carr}}, \bibinfo {author} {\bibfnamefont {M.}~\bibnamefont {Raidal}},
  \bibinfo {author} {\bibfnamefont {T.}~\bibnamefont {Tenkanen}}, \bibinfo
  {author} {\bibfnamefont {V.}~\bibnamefont {Vaskonen}}, \ and\ \bibinfo
  {author} {\bibfnamefont {H.}~\bibnamefont {Veermäe}},\ }\href {\doibase
  10.1103/PhysRevD.96.023514} {\bibfield  {journal} {\bibinfo  {journal} {Phys.
  Rev.}\ }\textbf {\bibinfo {volume} {D96}},\ \bibinfo {pages} {023514}
  (\bibinfo {year} {2017})},\ \Eprint {http://arxiv.org/abs/1705.05567}
  {arXiv:1705.05567 [astro-ph.CO]} \BibitemShut {NoStop}%
\bibitem [{\citenamefont {Garcia-Bellido}\ and\ \citenamefont
  {Ruiz~Morales}(2017)}]{Garcia-Bellido:2017mdw}%
  \BibitemOpen
  \bibfield  {author} {\bibinfo {author} {\bibfnamefont {J.}~\bibnamefont
  {Garcia-Bellido}}\ and\ \bibinfo {author} {\bibfnamefont {E.}~\bibnamefont
  {Ruiz~Morales}},\ }\href@noop {} {\  (\bibinfo {year} {2017})},\ \Eprint
  {http://arxiv.org/abs/1702.03901} {arXiv:1702.03901 [astro-ph.CO]}
  \BibitemShut {NoStop}%
\bibitem [{\citenamefont {Ezquiaga}\ \emph {et~al.}(2017)\citenamefont
  {Ezquiaga}, \citenamefont {Garcia-Bellido},\ and\ \citenamefont
  {Ruiz~Morales}}]{Ezquiaga:2017fvi}%
  \BibitemOpen
  \bibfield  {author} {\bibinfo {author} {\bibfnamefont {J.~M.}\ \bibnamefont
  {Ezquiaga}}, \bibinfo {author} {\bibfnamefont {J.}~\bibnamefont
  {Garcia-Bellido}}, \ and\ \bibinfo {author} {\bibfnamefont {E.}~\bibnamefont
  {Ruiz~Morales}},\ }\href@noop {} {\  (\bibinfo {year} {2017})},\ \Eprint
  {http://arxiv.org/abs/1705.04861} {arXiv:1705.04861 [astro-ph.CO]}
  \BibitemShut {NoStop}%
\bibitem [{\citenamefont {Drees}\ and\ \citenamefont
  {Erfani}(2011)}]{Drees:2011hb}%
  \BibitemOpen
  \bibfield  {author} {\bibinfo {author} {\bibfnamefont {M.}~\bibnamefont
  {Drees}}\ and\ \bibinfo {author} {\bibfnamefont {E.}~\bibnamefont {Erfani}},\
  }\href {\doibase 10.1088/1475-7516/2011/04/005} {\bibfield  {journal}
  {\bibinfo  {journal} {JCAP}\ }\textbf {\bibinfo {volume} {1104}},\ \bibinfo
  {pages} {005} (\bibinfo {year} {2011})},\ \Eprint
  {http://arxiv.org/abs/1102.2340} {arXiv:1102.2340 [hep-ph]} \BibitemShut
  {NoStop}%
\bibitem [{\citenamefont {Motohashi}\ and\ \citenamefont
  {Hu}(2015)}]{Motohashi:2015hpa}%
  \BibitemOpen
  \bibfield  {author} {\bibinfo {author} {\bibfnamefont {H.}~\bibnamefont
  {Motohashi}}\ and\ \bibinfo {author} {\bibfnamefont {W.}~\bibnamefont {Hu}},\
  }\href {\doibase 10.1103/PhysRevD.92.043501} {\bibfield  {journal} {\bibinfo
  {journal} {Phys. Rev.}\ }\textbf {\bibinfo {volume} {D92}},\ \bibinfo {pages}
  {043501} (\bibinfo {year} {2015})},\ \Eprint
  {http://arxiv.org/abs/1503.04810} {arXiv:1503.04810 [astro-ph.CO]}
  \BibitemShut {NoStop}%
\bibitem [{\citenamefont {Motohashi}\ and\ \citenamefont
  {Hu}(2017)}]{Motohashi:2017gqb}%
  \BibitemOpen
  \bibfield  {author} {\bibinfo {author} {\bibfnamefont {H.}~\bibnamefont
  {Motohashi}}\ and\ \bibinfo {author} {\bibfnamefont {W.}~\bibnamefont {Hu}},\
  }\href {\doibase 10.1103/PhysRevD.96.023502} {\bibfield  {journal} {\bibinfo
  {journal} {Phys. Rev.}\ }\textbf {\bibinfo {volume} {D96}},\ \bibinfo {pages}
  {023502} (\bibinfo {year} {2017})},\ \Eprint
  {http://arxiv.org/abs/1704.01128} {arXiv:1704.01128 [hep-th]} \BibitemShut
  {NoStop}%
\bibitem [{\citenamefont {Harada}\ \emph {et~al.}(2013)\citenamefont {Harada},
  \citenamefont {Yoo},\ and\ \citenamefont {Kohri}}]{Harada:2013epa}%
  \BibitemOpen
  \bibfield  {author} {\bibinfo {author} {\bibfnamefont {T.}~\bibnamefont
  {Harada}}, \bibinfo {author} {\bibfnamefont {C.-M.}\ \bibnamefont {Yoo}}, \
  and\ \bibinfo {author} {\bibfnamefont {K.}~\bibnamefont {Kohri}},\ }\href
  {\doibase 10.1103/PhysRevD.88.084051, 10.1103/PhysRevD.89.029903} {\bibfield
  {journal} {\bibinfo  {journal} {Phys. Rev.}\ }\textbf {\bibinfo {volume}
  {D88}},\ \bibinfo {pages} {084051} (\bibinfo {year} {2013})},\ \bibinfo
  {note} {[Erratum: Phys. Rev.D89,no.2,029903(2014)]},\ \Eprint
  {http://arxiv.org/abs/1309.4201} {arXiv:1309.4201 [astro-ph.CO]} \BibitemShut
  {NoStop}%
\bibitem [{\citenamefont {Young}\ \emph {et~al.}(2014)\citenamefont {Young},
  \citenamefont {Byrnes},\ and\ \citenamefont {Sasaki}}]{Young:2014ana}%
  \BibitemOpen
  \bibfield  {author} {\bibinfo {author} {\bibfnamefont {S.}~\bibnamefont
  {Young}}, \bibinfo {author} {\bibfnamefont {C.~T.}\ \bibnamefont {Byrnes}}, \
  and\ \bibinfo {author} {\bibfnamefont {M.}~\bibnamefont {Sasaki}},\ }\href
  {\doibase 10.1088/1475-7516/2014/07/045} {\bibfield  {journal} {\bibinfo
  {journal} {JCAP}\ }\textbf {\bibinfo {volume} {1407}},\ \bibinfo {pages}
  {045} (\bibinfo {year} {2014})},\ \Eprint {http://arxiv.org/abs/1405.7023}
  {arXiv:1405.7023 [gr-qc]} \BibitemShut {NoStop}%
\bibitem [{\citenamefont {Musco}\ and\ \citenamefont
  {Miller}(2013)}]{Musco:2012au}%
  \BibitemOpen
  \bibfield  {author} {\bibinfo {author} {\bibfnamefont {I.}~\bibnamefont
  {Musco}}\ and\ \bibinfo {author} {\bibfnamefont {J.~C.}\ \bibnamefont
  {Miller}},\ }\href {\doibase 10.1088/0264-9381/30/14/145009} {\bibfield
  {journal} {\bibinfo  {journal} {Class. Quant. Grav.}\ }\textbf {\bibinfo
  {volume} {30}},\ \bibinfo {pages} {145009} (\bibinfo {year} {2013})},\
  \Eprint {http://arxiv.org/abs/1201.2379} {arXiv:1201.2379 [gr-qc]}
  \BibitemShut {NoStop}%
\bibitem [{\citenamefont {Ade}\ \emph {et~al.}(2016{\natexlab{a}})\citenamefont
  {Ade} \emph {et~al.}}]{Ade:2015xua}%
  \BibitemOpen
  \bibfield  {author} {\bibinfo {author} {\bibfnamefont {P.~A.~R.}\
  \bibnamefont {Ade}} \emph {et~al.} (\bibinfo {collaboration} {Planck}),\
  }\href {\doibase 10.1051/0004-6361/201525830} {\bibfield  {journal} {\bibinfo
   {journal} {Astron. Astrophys.}\ }\textbf {\bibinfo {volume} {594}},\
  \bibinfo {pages} {A13} (\bibinfo {year} {2016}{\natexlab{a}})},\ \Eprint
  {http://arxiv.org/abs/1502.01589} {arXiv:1502.01589 [astro-ph.CO]}
  \BibitemShut {NoStop}%
\bibitem [{\citenamefont {Kinney}(2005)}]{Kinney:2005vj}%
  \BibitemOpen
  \bibfield  {author} {\bibinfo {author} {\bibfnamefont {W.~H.}\ \bibnamefont
  {Kinney}},\ }\href {\doibase 10.1103/PhysRevD.72.023515} {\bibfield
  {journal} {\bibinfo  {journal} {Phys. Rev.}\ }\textbf {\bibinfo {volume}
  {D72}},\ \bibinfo {pages} {023515} (\bibinfo {year} {2005})},\ \Eprint
  {http://arxiv.org/abs/gr-qc/0503017} {arXiv:gr-qc/0503017 [gr-qc]}
  \BibitemShut {NoStop}%
\bibitem [{\citenamefont {Dvorkin}\ and\ \citenamefont
  {Hu}(2010)}]{Dvorkin:2009ne}%
  \BibitemOpen
  \bibfield  {author} {\bibinfo {author} {\bibfnamefont {C.}~\bibnamefont
  {Dvorkin}}\ and\ \bibinfo {author} {\bibfnamefont {W.}~\bibnamefont {Hu}},\
  }\href {\doibase 10.1103/PhysRevD.81.023518} {\bibfield  {journal} {\bibinfo
  {journal} {Phys. Rev.}\ }\textbf {\bibinfo {volume} {D81}},\ \bibinfo {pages}
  {023518} (\bibinfo {year} {2010})},\ \Eprint {http://arxiv.org/abs/0910.2237}
  {arXiv:0910.2237 [astro-ph.CO]} \BibitemShut {NoStop}%
\bibitem [{\citenamefont {Namjoo}\ \emph {et~al.}(2013)\citenamefont {Namjoo},
  \citenamefont {Firouzjahi},\ and\ \citenamefont {Sasaki}}]{Namjoo:2012aa}%
  \BibitemOpen
  \bibfield  {author} {\bibinfo {author} {\bibfnamefont {M.~H.}\ \bibnamefont
  {Namjoo}}, \bibinfo {author} {\bibfnamefont {H.}~\bibnamefont {Firouzjahi}},
  \ and\ \bibinfo {author} {\bibfnamefont {M.}~\bibnamefont {Sasaki}},\ }\href
  {\doibase 10.1209/0295-5075/101/39001} {\bibfield  {journal} {\bibinfo
  {journal} {Europhys. Lett.}\ }\textbf {\bibinfo {volume} {101}},\ \bibinfo
  {pages} {39001} (\bibinfo {year} {2013})},\ \Eprint
  {http://arxiv.org/abs/1210.3692} {arXiv:1210.3692 [astro-ph.CO]} \BibitemShut
  {NoStop}%
\bibitem [{\citenamefont {Martin}\ \emph {et~al.}(2013)\citenamefont {Martin},
  \citenamefont {Motohashi},\ and\ \citenamefont {Suyama}}]{Martin:2012pe}%
  \BibitemOpen
  \bibfield  {author} {\bibinfo {author} {\bibfnamefont {J.}~\bibnamefont
  {Martin}}, \bibinfo {author} {\bibfnamefont {H.}~\bibnamefont {Motohashi}}, \
  and\ \bibinfo {author} {\bibfnamefont {T.}~\bibnamefont {Suyama}},\ }\href
  {\doibase 10.1103/PhysRevD.87.023514} {\bibfield  {journal} {\bibinfo
  {journal} {Phys. Rev.}\ }\textbf {\bibinfo {volume} {D87}},\ \bibinfo {pages}
  {023514} (\bibinfo {year} {2013})},\ \Eprint {http://arxiv.org/abs/1211.0083}
  {arXiv:1211.0083 [astro-ph.CO]} \BibitemShut {NoStop}%
\bibitem [{\citenamefont {Motohashi}\ \emph {et~al.}(2015)\citenamefont
  {Motohashi}, \citenamefont {Starobinsky},\ and\ \citenamefont
  {Yokoyama}}]{Motohashi:2014ppa}%
  \BibitemOpen
  \bibfield  {author} {\bibinfo {author} {\bibfnamefont {H.}~\bibnamefont
  {Motohashi}}, \bibinfo {author} {\bibfnamefont {A.~A.}\ \bibnamefont
  {Starobinsky}}, \ and\ \bibinfo {author} {\bibfnamefont {J.}~\bibnamefont
  {Yokoyama}},\ }\href {\doibase 10.1088/1475-7516/2015/09/018} {\bibfield
  {journal} {\bibinfo  {journal} {JCAP}\ }\textbf {\bibinfo {volume} {1509}},\
  \bibinfo {pages} {018} (\bibinfo {year} {2015})},\ \Eprint
  {http://arxiv.org/abs/1411.5021} {arXiv:1411.5021 [astro-ph.CO]} \BibitemShut
  {NoStop}%
\bibitem [{\citenamefont {Motohashi}\ and\ \citenamefont
  {Starobinsky}(2017{\natexlab{a}})}]{Motohashi:2017aob}%
  \BibitemOpen
  \bibfield  {author} {\bibinfo {author} {\bibfnamefont {H.}~\bibnamefont
  {Motohashi}}\ and\ \bibinfo {author} {\bibfnamefont {A.~A.}\ \bibnamefont
  {Starobinsky}},\ }\href {\doibase 10.1209/0295-5075/117/39001} {\bibfield
  {journal} {\bibinfo  {journal} {Europhys. Lett.}\ }\textbf {\bibinfo {volume}
  {117}},\ \bibinfo {pages} {39001} (\bibinfo {year} {2017}{\natexlab{a}})},\
  \Eprint {http://arxiv.org/abs/1702.05847} {arXiv:1702.05847 [astro-ph.CO]}
  \BibitemShut {NoStop}%
\bibitem [{\citenamefont {Motohashi}\ and\ \citenamefont
  {Starobinsky}(2017{\natexlab{b}})}]{Motohashi:2017vdc}%
  \BibitemOpen
  \bibfield  {author} {\bibinfo {author} {\bibfnamefont {H.}~\bibnamefont
  {Motohashi}}\ and\ \bibinfo {author} {\bibfnamefont {A.~A.}\ \bibnamefont
  {Starobinsky}},\ }\href {\doibase 10.1140/epjc/s10052-017-5109-x} {\bibfield
  {journal} {\bibinfo  {journal} {Eur. Phys. J.}\ }\textbf {\bibinfo {volume}
  {C77}},\ \bibinfo {pages} {538} (\bibinfo {year} {2017}{\natexlab{b}})},\
  \Eprint {http://arxiv.org/abs/1704.08188} {arXiv:1704.08188 [astro-ph.CO]}
  \BibitemShut {NoStop}%
\bibitem [{\citenamefont {Ade}\ \emph {et~al.}(2016{\natexlab{b}})\citenamefont
  {Ade} \emph {et~al.}}]{Ade:2015lrj}%
  \BibitemOpen
  \bibfield  {author} {\bibinfo {author} {\bibfnamefont {P.~A.~R.}\
  \bibnamefont {Ade}} \emph {et~al.} (\bibinfo {collaboration} {Planck}),\
  }\href {\doibase 10.1051/0004-6361/201525898} {\bibfield  {journal} {\bibinfo
   {journal} {Astron. Astrophys.}\ }\textbf {\bibinfo {volume} {594}},\
  \bibinfo {pages} {A20} (\bibinfo {year} {2016}{\natexlab{b}})},\ \Eprint
  {http://arxiv.org/abs/1502.02114} {arXiv:1502.02114 [astro-ph.CO]}
  \BibitemShut {NoStop}%
\bibitem [{\citenamefont {Kannike}\ \emph {et~al.}(2017)\citenamefont
  {Kannike}, \citenamefont {Marzola}, \citenamefont {Raidal},\ and\
  \citenamefont {Veermäe}}]{Kannike:2017bxn}%
  \BibitemOpen
  \bibfield  {author} {\bibinfo {author} {\bibfnamefont {K.}~\bibnamefont
  {Kannike}}, \bibinfo {author} {\bibfnamefont {L.}~\bibnamefont {Marzola}},
  \bibinfo {author} {\bibfnamefont {M.}~\bibnamefont {Raidal}}, \ and\ \bibinfo
  {author} {\bibfnamefont {H.}~\bibnamefont {Veermäe}},\ }\href@noop {} {\
  (\bibinfo {year} {2017})},\ \Eprint {http://arxiv.org/abs/1705.06225}
  {arXiv:1705.06225 [astro-ph.CO]} \BibitemShut {NoStop}%
\bibitem [{\citenamefont {Germani}\ and\ \citenamefont
  {Prokopec}(2017)}]{Germani:2017bcs}%
  \BibitemOpen
  \bibfield  {author} {\bibinfo {author} {\bibfnamefont {C.}~\bibnamefont
  {Germani}}\ and\ \bibinfo {author} {\bibfnamefont {T.}~\bibnamefont
  {Prokopec}},\ }\href@noop {} {\  (\bibinfo {year} {2017})},\ \Eprint
  {http://arxiv.org/abs/1706.04226} {arXiv:1706.04226 [astro-ph.CO]}
  \BibitemShut {NoStop}%
\end{thebibliography}%

\end{document}